\documentclass[10pt, conference]{IEEEtran}
\IEEEoverridecommandlockouts
\usepackage{cite}
\usepackage{verbatim}
\usepackage{url}
\usepackage{grffile}
\usepackage{graphicx}
\usepackage{subcaption}
\usepackage{amsmath,amssymb,amsfonts}
\usepackage[detect-all]{siunitx}
\usepackage[hidelinks]{hyperref}
\usepackage{cleveref}
\usepackage{tikz}
\usepackage{algorithmic}
\usepackage{epstopdf}
\usepackage[utf8]{inputenc}

\def\BibTeX{{\rm B\kern-.05em{\sc i\kern-.025em b}\kern-.08em
    T\kern-.1667em\lower.7ex\hbox{E}\kern-.125emX}}

\newcommand\copyrighttext{%
  \footnotesize \textcopyright 2024 IEEE. Personal use of this material is permitted. Permission from IEEE must be obtained for all other uses, in any current or future media, including reprinting/republishing this material for advertising or promotional purposes, creating new collective works, for resale or redistribution to servers or lists, or reuse of any copyrighted component of this work in other works.}

\newcommand\copyrightnotice{%
\begin{tikzpicture}[remember picture,overlay]
\node[anchor=south,yshift=10pt] at (current page.south) {\fbox{\parbox{\dimexpr\textwidth-\fboxsep-\fboxrule\relax}{\copyrighttext}}};
\end{tikzpicture}%
}    

\usepackage[]{footmisc}

\begin{document}

\title{Conference Paper Title*\\
{\footnotesize \textsuperscript{*}Note: Sub-titles are not captured in Xplore and
should not be used}
\thanks{Identify applicable funding agency here. If none, delete this.}
}

\title{A Mobility Analysis of UE-Side Beamforming for Multi-Panel User Equipment with Hand Blockage  \vspace{-0.2\baselineskip}\\
}

\author{
    \IEEEauthorblockN{
        Subhyal Bin Iqbal\IEEEauthorrefmark{1}\IEEEauthorrefmark{2}, Salman Nadaf\IEEEauthorrefmark{1}\IEEEauthorrefmark{3},  Umur Karabulut\IEEEauthorrefmark{1}, Philipp Schulz\IEEEauthorrefmark{2}, Anna Prado\IEEEauthorrefmark{3}, \\ Gerhard P. Fettweis\IEEEauthorrefmark{2} and Wolfgang Kellerer \IEEEauthorrefmark{3} 
    }

    \IEEEauthorblockA{\IEEEauthorrefmark{1} Nokia Standardization and Research Lab, Munich, Germany}
    \IEEEauthorblockA{\IEEEauthorrefmark{2} Vodafone Chair for Mobile Communications Systems, Technische Universität Dresden, Germany}
    \IEEEauthorblockA{\IEEEauthorrefmark{3} Chair of Communication Networks, Technical University of Munich, Germany}
\vspace{-2\baselineskip}
}

\maketitle

%
\copyrightnotice

\begin{abstract}
The hand blockage effect of the human hand around the user equipment (UE) is too considerable to be ignored in frequency range 2 (FR2). This adds another layer of complexity to the link budget design in FR2 for 5G networks, which already suffer from high path and diffraction loss. More recently, multi-panel UEs (MPUEs) have been proposed as a way to address this problem, whereby multiple distinct antenna panels are integrated into the UE body as a way to leverage gains from antenna directivity. MPUEs also enhance the Rx-beamforming gain because it is now subject to each individual antenna panel. In this paper, the mobility performance of hand blockage induced by three practical hand grips is analyzed in a system-level simulation, where in each grip both the UE orientation and the hand positioning around the UE is different. It is seen that each hand grip has a significant impact on mobility performance of the network, where in the worst case mobility failures increase by 43\% compared to the non-hand blockage case. Moreover, a detailed analysis of the tradeoff between the mobility key performance indicators and the panel and Rx beam switching frequency is also studied. Results have shown that both the panel and Rx beam switches can be reduced considerably without compromising on the mobility performance. This is beneficial because it helps in reducing UE power consumption. 

\end{abstract}


\section{Introduction} \label{Section1}
Although the 5th generation (5G) of mobile networks relies heavily on the large spectrum availability offered by frequency range 2 (FR2) \cite{b1}, it introduces further challenges to radio propagation such as higher free-space path loss and penetration loss \cite{b2}. To improve the link budget through beamforming gain, narrow beams are integrated into the network, both on the next-generation node base station (gNb) and the user equipment (UE). Given the directional nature of these antennas and the unpredictable UE orientation which is subject to how humans use and grip these devices in everyday life, the introduction of multi-panel UEs (MPUEs) \cite{b3, b4} is one way to realize the improvement in the network dimensioning that is needed. However, it is also well-known that at these higher frequencies in the order of 28 GHz, the penetration depth into the human hand holding the UE is very small \cite{b5, b6}. This results in a higher degree of blockage by the hand and also needs to be taken into account in the link budget design.

In our earlier study\cite{b7}, we investigated the system-level mobility performance of MPUEs with UE-side receiver (Rx)-beamforming that individually caters towards improving both the serving and target links in order to reduce mobility failures. Furthermore, the work also focused on making the handover (HO) process more reliable by integrating the Rx-beamformed measurements into the HO process. The Rx-beamforming framework is based on 3GPP’s hierarchical beamforming framework \cite[Section 6.1]{b8}. The mobility performance was analyzed in terms of an extended mobility key performance indicator (KPI) set pertaining to both inter-cell and intra-cell mobility. In another associated work \cite{b5}, the authors have studied the mobility performance of a network with beamforming both at the gNb and UE-side together with the hand blockage impact for MPUEs. However, this study was limited to intra-cell mobility and did not focus on the handovers and mobility failures, both of which are essential KPIs to quantify the mobility performance in a network. As such, a study is still missing on investigating both the intra and the inter-cell mobility performance of MPUE that incorporates UE-side Rx-beamforming in conjunction with hand blockage. This paper aims to bridge the gap by analyzing the impact of different hand grips and their associated hand blockage in a 5G network where both Tx and Rx beamforming are employed. As such, this is its first major contribution. 


One of the main impediments to the development of effective FR2 communication devices for mobile applications is UE power consumption. In this study, we differentiate between steady-sate UE power consumption and switching power consumption, where the latter is associated with panel and Rx beam switches. The authors in \cite{b9} have provided a detailed analysis of the steady-state power consumption of different beamforming methods such as analog, digital, and hybrid beamforming. However, a power consumption analysis for the panel and Rx beam switching is not performed. While the authors in \cite{b3} do introduce panel switching offsets and analyze the mobility performance of MPUEs and the associated panel switching frequency, the study is not extended to Rx-beamforming capable MPUEs and does not include hand blockages. In the second contribution of our paper, the impact of the panel and Rx beam switching offsets on both the mobility performance and the switching frequency is studied.


\section{Network Model} \label{Section2}
In this section, the inter-cell HO and beam management procedures, including both Tx and UE-side Rx-beamforming procedures are reviewed. The panel and Rx beam switching mechanism is then later explained.

\subsection{HO Model} \label{Subection2.1}
HOs between different cells in the network attribute to inter-cell mobility. For a successful HO from the serving cell to the target cell, a pre-requisite is the filtering of physical layer reference signal received power (RSRP) measurements to mitigate the effect of channel impairments. When a 5G beamformed network is considered, each UE is assumed to be capable of measuring the raw RSRP values $P_{c,b}^\textrm{RSRP}(n)$ at a discrete time instant $n$ from each Tx beam $b \in B$ of cell $c \in C$, using the synchronization signal block (SSB) bursts that are periodically transmitted by the base station (BS). The separation between the time instants is denoted by $\Delta t$ ms. At the UE side, layer 1 (L1) and L3 filtering are then sequentially applied to the raw RSRPs in order to counter the effects of fast fading and measurement errors and determine the L3 cell quality of the serving and neighboring cells. A more detailed explanation of the L1 and L3 filtering procedures can be found in \cite[Section II.A]{b7}. In this paper, the HO model is based on the baseline HO mechanism defined in \textit{3GPP Release 15} \cite{b1}.

L3 cell quality $P_{c}^\textrm{L3}(m)$ is an important indicator of the average downlink signal strength for a link that exists between a UE and cell~$c$. The network uses the L3 measurements to trigger an HO from the serving cell $c_0$ to one of its neighboring cells, termed as the target cell $c^{\prime}$. For intra-frequency HO decisions, generally the A3 trigger condition is configured for measurement reporting~\cite{b1}. The UE is configured to report the L3 cell quality measurement $P_{c^{\prime}}^\textrm{L3}(m)$ and L3 beam measurements $P_{c^{\prime},b}^\textrm{L3}(m)$ of the target cell to its serving cell $c_0$ when the A3 trigger condition, i.e.,
\vspace{-0.2\baselineskip}
\begin{equation}
\label{Eq1}
     P_{c_0}^\textrm{L3}(m) + o^\mathrm{A3}_{c_0, c^{\prime}} < P_{c^{\prime}}^\textrm{L3}(m) \ \text{for} \  m_0 - T_\mathrm{TTT,A3} < m < m_0,
\end{equation}
expires at the time instant $m=m_0$ for $c^{\prime} \neq c_0$, where $o^\mathrm{A3}_{c_0, c^{\prime}}$ is termed as the HO offset between cell $c_0$ and $c^{\prime}$ and the observation period in (\ref{Eq1}) is termed as the time-to-trigger $T_\mathrm{TTT,A3}$ (in ms).

After receiving the L3 measurements from the UE, the serving cell $c_0$ sends out an HO request to the target cell~$c^{\prime}$, which is typically the strongest cell in the network, along with the L3 beam measurements $P_{c^{\prime},b}^\textrm{L3}(m)$ of $c^{\prime}$. The target cell then prepares contention free random-access (CFRA) resources for beams $b \in B_{\mathrm{prep},c^{\prime}}$ with the highest signal power based on the L3 beam measurements that have been reported. The target cell thereafter provides an HO request acknowledgement to the serving cell, which thereupon sends an HO command to the UE. Once the UE receives this message, it detaches from its serving cell $c_0$ and initiates random-access towards the target cell $c^{\prime}$ using these CFRA resources.

\subsection{Tx Beamforming} \label{Subection2.2}
In a 5G beamformed network, typically beamforming is employed both on the gnB side and the UE side. Beamforming on the Tx-side attributes to intra-cell mobility, where the cells controlled by the gnB in the network are capable of generating multiple directional beams along different azimuth and elevation angles. In particular, it refers to a set of L1 and L2 Tx beam management procedures for the determination and update of serving Tx beam(s) for each UE within a serving cell $c_0$, as defined in \textit{3GPP Release 15} \cite{b8}. This includes Tx beam selection, where the UE uses network assistance to choose the serving beam $b_0$ that it thereafter uses to communicate with $c_0$ \cite[Section II-B]{b4}. The other key component is beam failure detection, where the UE monitors the radio link quality signal-to-noise-plus-interference ratio (SINR) \cite{b10} and detects a failure of the serving beam when the SINR falls below a certain threshold. In this case, the UE initiates a beam failure recovery procedure where it tries to recover another beam of the serving cell $c_0$. To that end, the UE performs a random access attempt on the target beam $b^{\prime}$ that has the highest L1 RSRP beam measurement $P_{c_0,b}^\textrm{L1}(m)$ value and then waits for the BS to send a random access response indicating that the access was successful. In the event of an unsuccessful attempt, the UE tries another random access using $b^{\prime}$. In total, $N_\mathrm{BAtt}$ such attempts are made at time intervals of $T_\mathrm{BAtt}$. If all such attempts are unsuccessful, a radio link failure (RLF) is declared by the UE.

\subsection{UE-side Rx-beamforming} \label{Subection2.3}
On the UE side, an MPUE in \textit{edge} design with three integrated panels is considered \cite{b3, b4}. A 1$\times$4 configuration with a 0.5$\lambda$ spacing between the antenna elements is assumed for each of the directional panels $d \in D$, where the antenna element radiation pattern is based on \cite{b8}. Each panel is considered to have beamforming capabilities for Rx-beamforming refinement {b8}, generating directional beams $r \in R$ for each of its panels $d \in D$, where $r \in \{1,\ldots,7\}$. The MPUE design and the Rx beam configuration design is modeled using  \textit{Computer Simulation Technology (CST) Microwave Studio} \cite{b14}. 


In our earlier work \cite{b4}, the concept of serving and best panel for the MPUE architecture was introduced. The former is used for measurement reporting for Tx beamforming while the latter is used in the HO procedure. Following upon that, in our other study \cite{b7} the allied concept of the serving and best Rx beam was introduced in the context of Rx-beamforming. In line with 3GPP \cite{b11}, the signal measurement scheme that we consider in this study is MPUE-Assumption 3 (MPUE-A3), where it is assumed that the UE can measure the RSRP values from the serving cell $c_0$ and neighboring cells by simultaneously activating all of its three panels. In the context of hierarchical beamforming \cite[Section 6.1]{b8}, 3GPP has defined UE-side Rx-beamforming as a follow-up procedure to the Tx serving beam selection procedure discussed earlier in \Cref{Subection2.2}. Once the serving beam $b_0$ based on SSB transmission has been selected, the UE can sweep through its set of Rx beams and thereby select a narrow refined beam. Herein, the serving cell now repeats the channel state information reference signal (CSI-RS) associated with the serving beam $b_0$ for some time while the UE is sweeping its Rx beams on its panels. In our implementation, we make two key assumptions. It is assumed that the serving beam on CSI-RS and the serving beam based on SSB have the same beamwidth. Furthermore, it is assumed that the Rx beam sweep for $b_0$ can be completed within the designated SSB period. Upon completion of the Rx beam sweep, the UE selects the beam with the highest L1 RSRP. The serving panel~$d_0$ and serving Rx beam $r_0$ are defined as
\vspace{-0.25\baselineskip}
\begin{equation}
\label{Eq2} 
[d_0, r_0] = \arg \max_{d,r} P_{c_0,b_0,d,r}^\mathrm{L1}(m).
\end{equation}
The serving panel $d_0$ serves two purposes \cite{b4}. It is used for beam reporting for intra-cell Tx beam management, as discussed in \Cref{Subection2.2}. Secondly, the raw beam panel RSRPs corresponding to $[d_0, r_0]$ are used for calculating the SINR $\gamma_{c,b}$ of a link between the UE and beam $b$ of cell $c$, which is used in handover failure (HOF) and RLF determination.

The best Rx beam~$r_c$ is chosen as the beam with the strongest L1 beam panel RSRP on the best panel~$d_c$ for any beam~$b$ of cell~$c$ in the network and is defined as
\vspace{-0.4\baselineskip}
\begin{equation}
\label{Eq3} 
[d_c, r_c] = \arg \max_{b,d,r} P_{c,b,d,r}^{\mathrm{L1}}(m).
\end{equation}
Herein, it is assumed that the UE can determine the best Rx beam with respect to the strongest L1 beam panel measurement of cell $c$. The L1 beam panel RSRPs $P_{c,b,d,r}^{\mathrm{L1}}(m)$ of the best panel $d_c$ and best Rx beam $r_c$ are denoted as L1 beam RSRPs $P_{c,b}^{\mathrm{L1}}(m)$ and are used for deriving the L3 cell quality measurement $P_{c}^{\mathrm{L3}}(m)$ and L3 beam measurements $P_{c,b}^{\mathrm{L3}}(m)$. These L3 cell quality measurements are then used for inter-cell HO determination, as explained in \Cref{Subection2.1}. 

When UE-side Rx-beamforming is used with MPUEs, it opens different approaches whereby Rx-beamforming can be used in the system model. This study considers \textit{Approach 3} defined in \cite[Section III.C]{b7}. In this approach, Rx-beamforming improves the serving link and therefore impacts both the Tx beam management and the serving link SINR. It also incorporates the enhancement proposed in \cite{b12} of acquiring the refined Rx target beam before an HO, whereby HOFs can be reduced because the target link now also benefits from the Rx-beamforming gain. Additionally, \textit{Approach 3} also exploits an area yet open to standardization, i.e., whether or not Rx-beamformed measurements are used in HO decisions \cite{b13}. 


\subsection{Panel and Beam Switching} \label{Subection2.4}
The aim of having multiple antenna panels and Rx-beamforming is to use the best Tx-Rx beam combination for communication between the UE and serving cell $c_0$ through the serving panel $d_0$ that has been determined in  (\ref{Eq2}). However, this would lead to frequent switching of both the MPUE panels and the Rx-beams generated on a particular panel. For an MPUE, we assume a separate radio frequency (RF) chain for each panel. Hence, switching to another serving panel means activating that panel for communication with the serving cell and if this is frequent, too many RF chains are being activated consecutively. We also assume analog Rx-beamforming. This means for each panel, there is one RF chain and multiple phase shifters feeding the elements of the antenna panel, which is limited to one serving Rx beam at a time. In order to switch to a new Rx beam, the phase shift of the Rx antenna elements needs to be changed so that all signals add constructively in a specific direction \cite{b15}. Both switching instances drain the UE battery and adversely affect the UE power consumption.


In order to prevent such frequent switching, we introduce a serving panel switching offset $o^\mathrm{p}$. As such, the serving panel $d_0$ and serving Rx beam $r_0$ are now determined as 
\vspace{-0.25\baselineskip}
\begin{equation}
\label{Eq4} 
[d_0, r_0] = \arg \max_{d,r} \left(P_{c_0,b_0,d,r}^\mathrm{L1}(m) + o^\mathrm{p}_{d,r}\right),
\end{equation}
where $o^\mathrm{p}_{d,r}$ = $0$ for [$d=d_0, r=r_0$] and $o^\mathrm{p}_{d,r}$ = $o^\mathrm{p}$ otherwise.

Similarly, a serving Rx beam switching offset $o^\mathrm{b}$ is introduced such that the serving Rx beam is now determined~as 
\vspace{-0.25\baselineskip}
\begin{equation}
\label{Eq5} 
r_0 = \arg \max_{r} \left(P_{c_0,b_0,d_0,r}^\mathrm{L1}(m) + o^\mathrm{b}_{r}\right),
\end{equation}
where $o^\mathrm{b}$ = $0$ for $r=r_0$ and $o^\mathrm{b}_{r}$ = $o^\mathrm{r}$ otherwise.



\section{Hand Blockage in MPUEs} \label{Section3}
\begin{figure*}[!t]    \begin{subfigure}{0.30\textwidth}        
\centering        \includegraphics[width=1\textwidth]{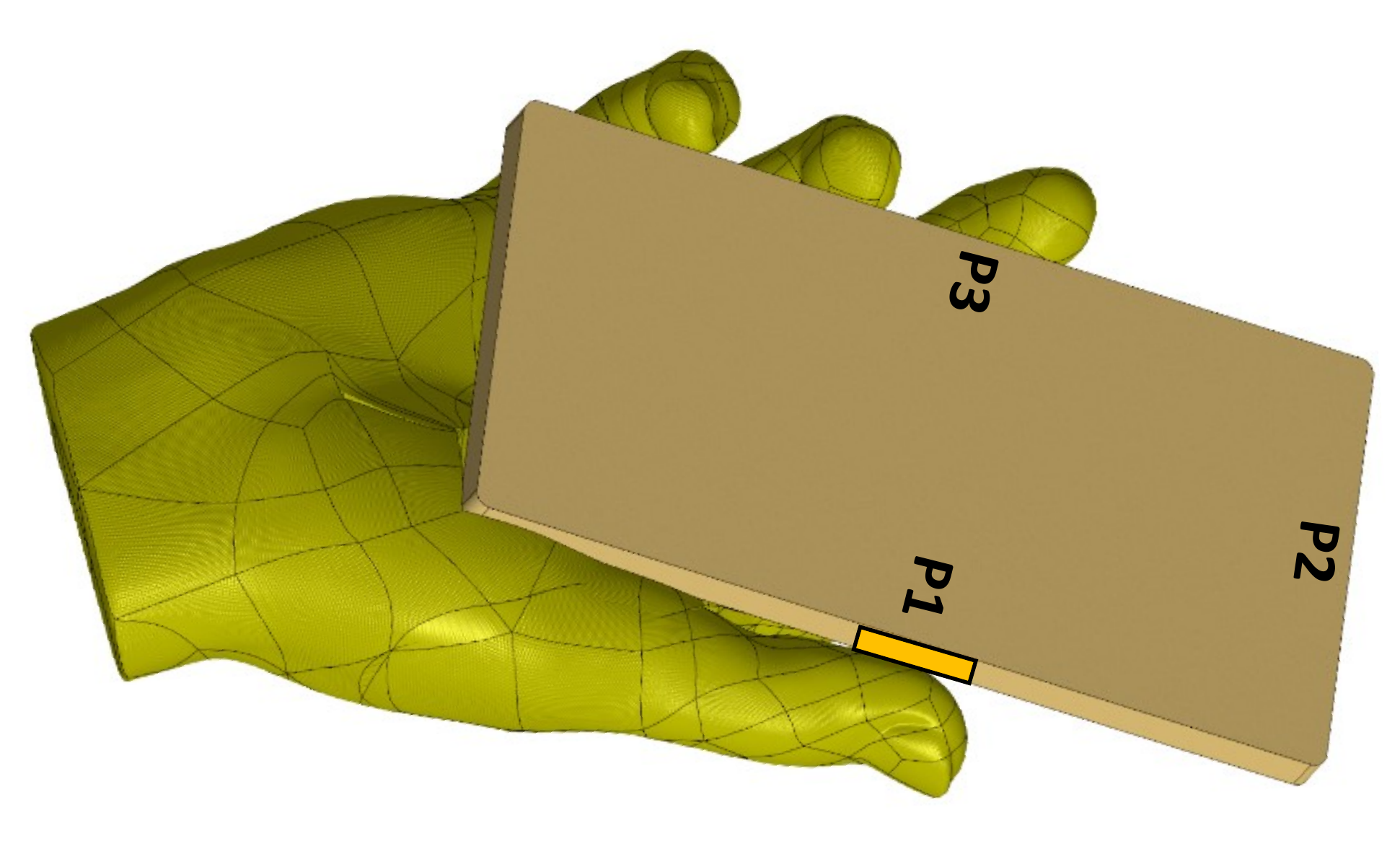}               \caption{Right-hand browsing grip (RHB).}        \label{fig:Fig2a}    \end{subfigure}\hfill    
\begin{subfigure}{0.28\textwidth}        
\centering        \includegraphics[width=1\textwidth]{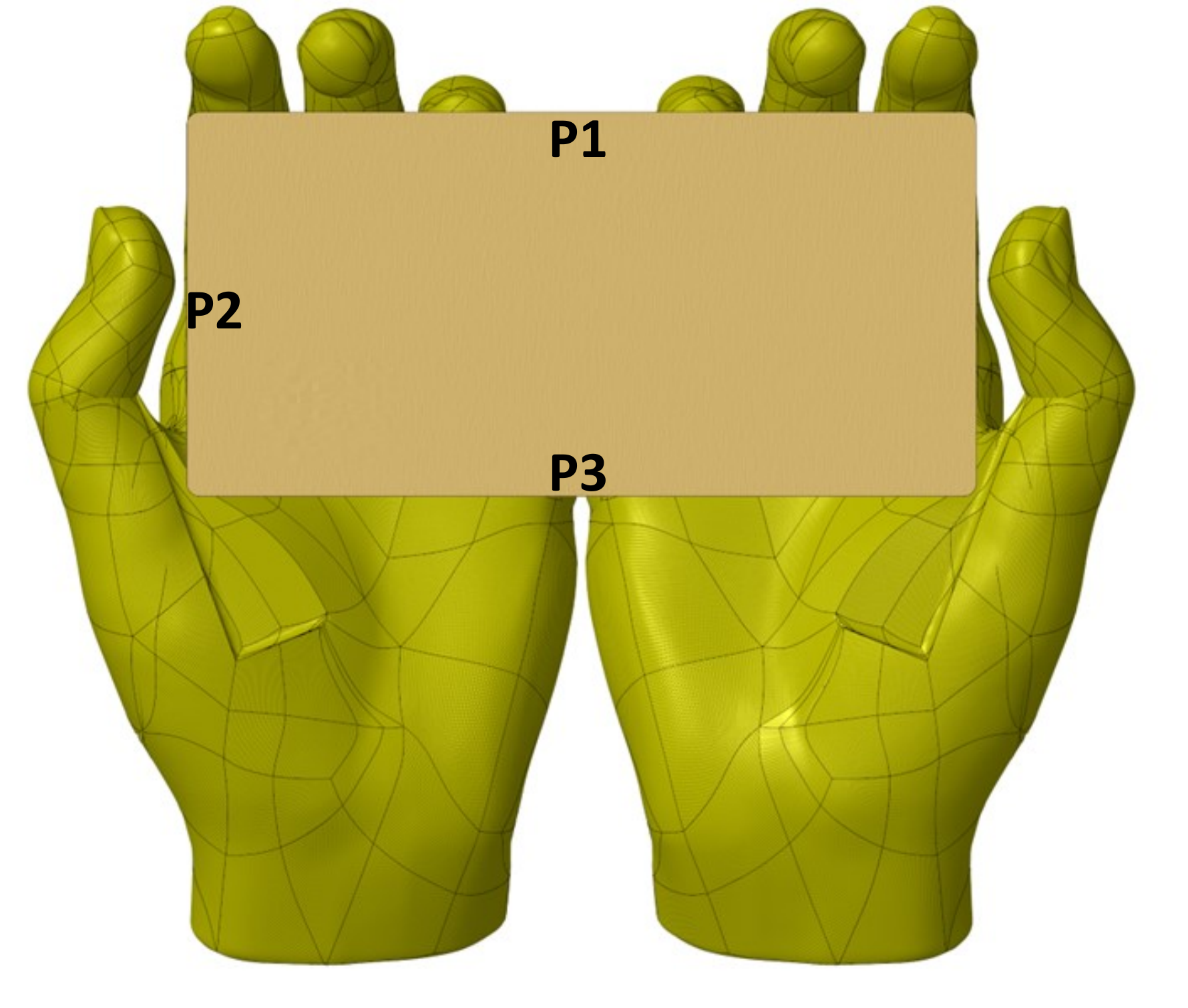}              \caption{Dual-hand streaming (DHS) grip.}        \label{fig:Fig2b}    \end{subfigure}\hfill 
\begin{subfigure}{0.36\textwidth}        
\centering        \includegraphics[width=1\textwidth]{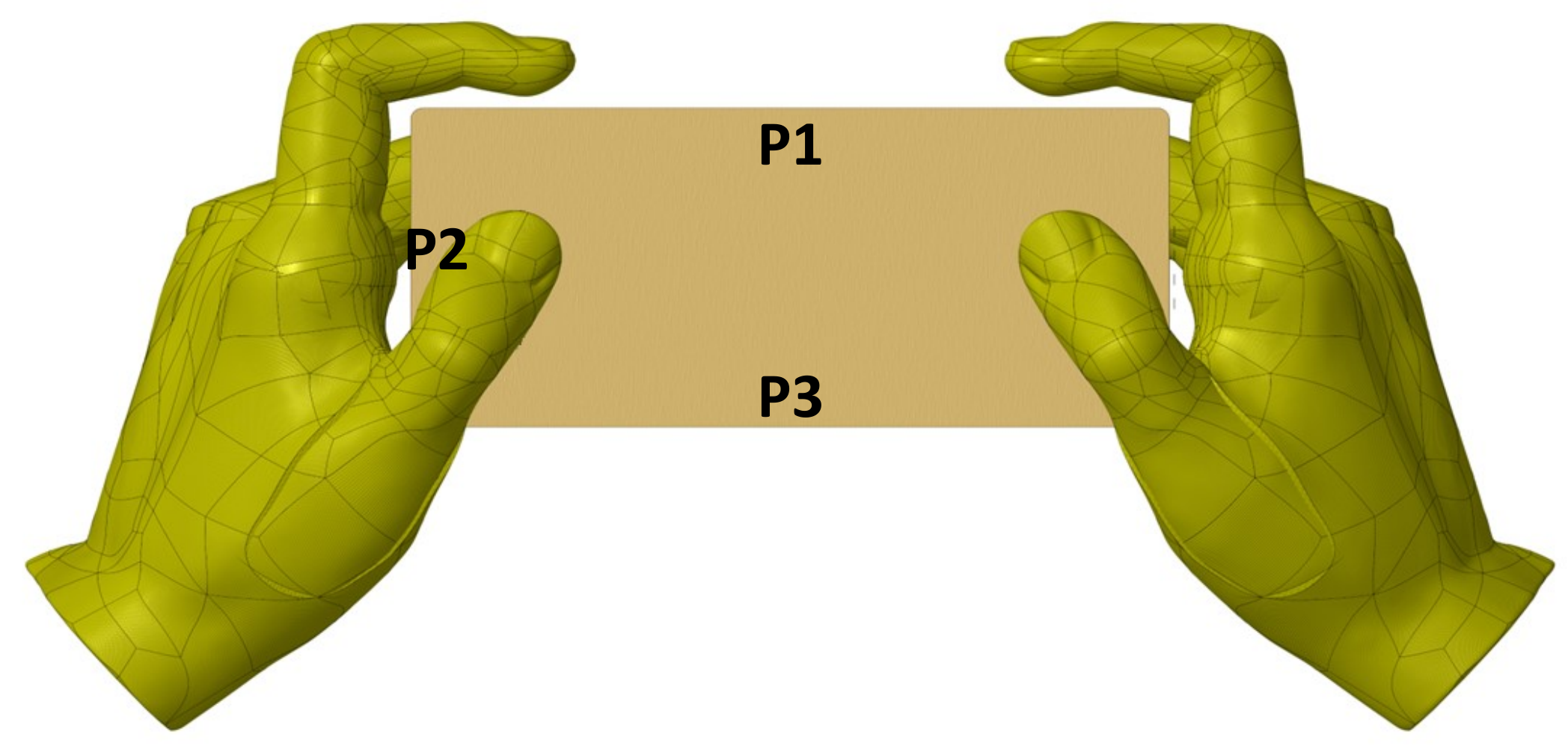}                \caption{Dual-hand gaming (DHG) grip.}        \label{fig:Fig2c}    \end{subfigure}\hfill 
\vspace{-1pt}
\caption{The three different hand grip models, which introduce a varying degree of hand blockage to the three different panels for an MPUE in the \textit{edge} design. The MPUE, hand blockage and the Rx-beams (not shown here) have been modeled using \textit{CST Microwave Studio} \cite{b14}.}     \label{fig:Fig2}
\vspace{-1.1\baselineskip}  \end{figure*}

In this work, three different hand grips for hand blockage based on \cite[Section III.B]{b5} are studied, comprising one portrait and two landscape orientations as shown in Fig.\,\ref{fig:Fig2}.

The right-hand browsing (RHB) grip is shown in Fig.\,\ref{fig:Fig2a} and is a representative of how humans normally use the phone when browsing or streaming content. It is different from hand the defined grip in \cite{b5} because now the UE is assumed parallel to the ground. As such, the use case changes from video calls to browsing or content streaming. It can be seen that the user's thumb is over panel 1 (P1) and the three fingers clench the device in close proximity to P3, with P2 avoiding any direct blockage by the hand. The dual-hand streaming (DHS) grip is shown in  Fig.\,\ref{fig:Fig2b} and mimics how usually humans stream content on their phones over prolonged periods, where the UE now needs to be held with both hands due to the large UE form factor sizes. Here, P2 is slightly covered by the user's right thumb, while P1 and P3 experience partial blockage from the user's fingers and palms, respectively. The dual-hand gaming (DHG) grip is shown in Fig.\,\ref{fig:Fig2c}. As the name implies, it models a gaming grip where the user is primarily interacting with the UE screen. It can be visualized that P2 is completely blocked by the left hand, while P1 and P3 remain uncovered. These hand grips are variations of Cellular Telecommunications and Internet Association's wide grip model \cite{b16} and have been modeled using \textit{CST Microwave Studio} \cite{b14}.



\section{Simulation Scenario and Parameters} \label{Section4}
In this section, the simulation setup for the 5G network model is explained. The simulations have been performed in our proprietary MATLAB-based system-level simulator, which incorporates the Rx-beamforming radiation patterns with the hand blockage effect, generated through \textit{CST Microwave Studio}, into the link budget design.

\begin{figure}[!t]
\textit{\centering
\includegraphics[width = 0.96\columnwidth]{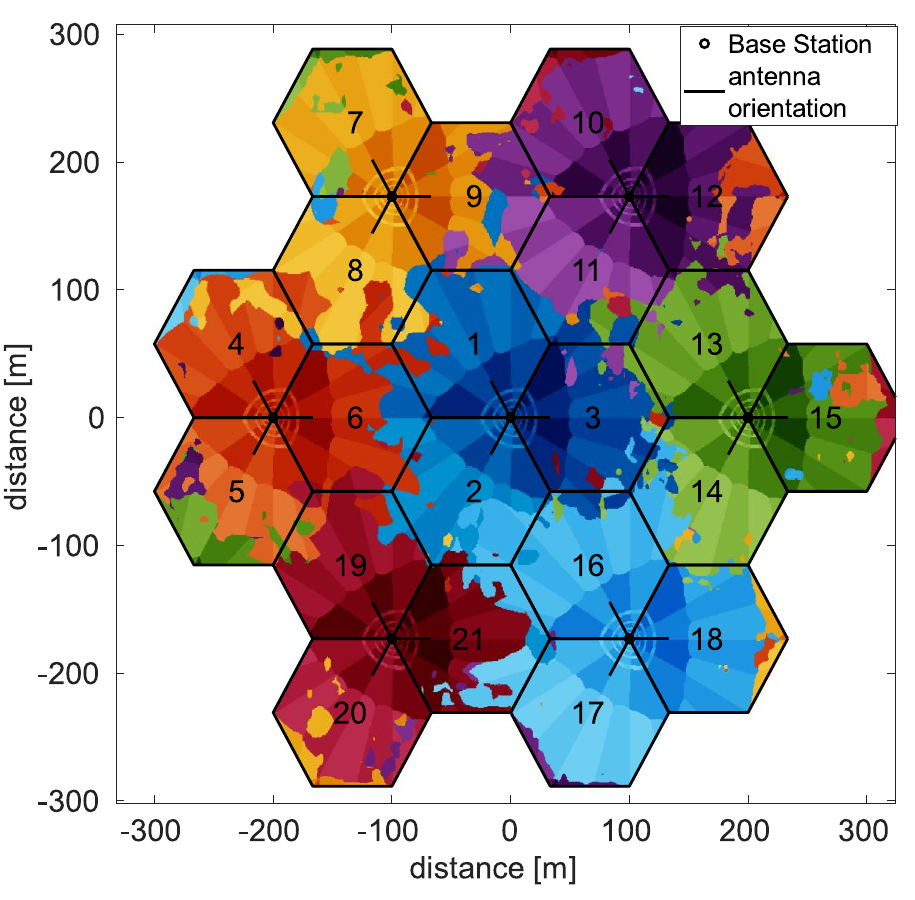}
\vspace{-7pt}
\caption{Simulation scenario consisting of seven hexagonal sites, where each site is serving three cells with 120$^{\circ}$ coverage. The effect of shadow fading
is also visible in the form of coverage islands.} 
\label{fig:Fig3}} 
\vspace{-1.2\baselineskip} \end{figure}

A network model with an urban-micro cellular deployment consisting of a standard hexagonal grid with seven BS sites is considered, each divided into three cells. The carrier frequency is 28 GHz and inter-cell distance is 200 m. At the start of the simulation, $N_\mathrm{UE}$ = 420 UEs are dropped randomly and follow a 2D uniform distribution over the network, moving at constant velocities along straight lines into random directions \cite[Table 7.8-5]{b1}. 
The UE speed is 60 km/h, which is the usual speed in the non-residential urban areas of cities~\cite{b7}. 
A wrap-around \cite[pp. 140]{b18} is considered, meaning that the hexagonal grid 
is repeated around the original 
layout shown in Fig.\,\ref{fig:Fig3} in the form of six replicas. The implication is that the no boundary effect occurs with respect to interference. The complete simulation parameters are listed in Table II in \cite[Section IV]{b4}.

As per 3GPP's study outlined in \textit {3GPP Release 15} \cite{b17}, the channel model takes into account shadow fading as  a result of large obstacles and assumes a soft line-of-sight (LoS) for all radio links between the cells and UEs. Soft LoS is defined as a weighted average of the LoS and non-LoS channel components \cite[pp. 59-60]{b17} and is used for both shadow fading and distance-dependent path loss determination. Fast fading is modeled through the low complexity channel model for multi-beam systems proposed in \cite{b20}, which integrates the spatial and temporal characteristics of 3GPP's geometry-based stochastic channel model (GSCM) \cite{b17} into Jakes’ channel model. The Tx-side beamforming gain model is based on \cite{b20}, where a 12-beam grid configuration is considered. $K_b=$ 4 beams are simultaneously scheduled  for all cells in the network. Beams $b \in \{1,\ldots,8\}$ have smaller beamwidth and higher beamforming gain and cover regions further apart from the BS. Beams  $b \in \{9,\ldots,12\}$ have larger beamwidth and relatively smaller beamforming gain and cover regions closer to the BS. This can be seen in Fig.\,\ref{fig:Fig3}, where the eight outer beams are shown in light color and the four inner beams are shown in dark color.

In order to evaluate the instantaneous downlink SINR $\gamma_{c,b}(m)$ of a link between the UE and beam $b$ of cell $c$ the Monte-Carlo approximation given in \cite{b21} for the strict fair resource scheduler is used, where all UEs are allotted the same amount of resources. This SINR is of key importance in the HOF and RLF models, both of which are based on the SINR threshold $\gamma_\mathrm{out} = \SI{-8}{dB}$. The former models failure of a UE to execute an HO from the serving cell $c_0$ to the prepared target cell $c^{\prime}$ and the latter models failure of a UE while still in cell $c_0$. They are explained in more detail in \cite[Section IV]{b4}.

\section{Performance Evaluation} \label{Section5}
In this section, the mobility performance of the three different hand grips is compared. Furthermore, the impact of the beam and panel switching offsets is also studied. The KPIs 
for both of these studies
are explained below.

\subsection{KPIs} \label{Subection5.1}






\textit{Successful HOs}: Sum of the total number of successful HOs from the serving  to the target cells in the network.

\textit{Mobility failures}: Sum of the total HOFs and RLFs in the network.

\textit{Fast HOs}: Sum of ping-pongs and short-stays in the network. A ping-pong is a successful HO followed by an HO back to the original cell within a very short time $T_\textrm{FH}$ \cite{b4}, e.g., 1~s. It is assumed that both these HOs could have been potentially avoided. A short-stay is a HO from one cell to another and then to a third one within $T_\textrm{FH}$. Here it is assumed that a direct HO from the first cell to the third one would have sufficed. Although fast HOs are still considered as successful HOs, they are accounted as a detrimental mobility KPI which adds outage and unnecessary signaling overhead to the network.    

\textit{Serving Panel Switches}: Sum of the serving panel switches for all UEs in the network. The serving panel switches due to HO or intra-cell serving Tx beam change are excluded in order to solely study the impact of the UE needing to switch to another serving panel.

\textit{Serving Rx Beam Switches}: Sum of the serving Rx beam switches for all UEs in the network. Excludes the serving Rx beam switches due to HO or serving panel change.  


The KPIs above are normalized to number of UEs $N_\mathrm{UE}$ in the network and to time and, thus, expressed as KPI/UE/min.


\textit{Outage:} Outage is defined as the time period when a UE is unable to receive data from the network. When the average downlink SINR of the serving cell $\gamma_{c_0, b_0}$ falls below $\gamma_\mathrm{out}$ it is assumed that the UE is unable to communicate with the network and, thus, in outage. Besides, if the HOF timer $T_{\mathrm{HOF}}$ expires due to a HOF or the RLF timer $T_{\mathrm{RLF}}$ expires due to an RLF, the UE initiates connection re-establishment \cite{b4} and this is also accounted for as outage. Despite being a necessary inter-cell mobility procedure, a successful HO also contributes to outage since the UE cannot receive any data during the time duration it is performing random access to the target cell $c^{\prime}$. This outage is modeled as relatively smaller (55~ms) than the outage due to connection re-establishment (180~ms) \cite{b4}. Outage is denoted in terms of a percentage 
\vspace{-1mm}
\begin{equation}
\label{Eq6} 
\textrm{Outage} \ (\%) = \frac{\sum_{u}{\textrm{Outage duration of UE}} \ u} {N_\mathrm{UE} \ \cdot \ \textrm{Simulated time}} \ \cdot \ 100. 
\end{equation}
%


\subsection{Simulation Results} \label{Subection5.2}
\begin{figure}[!t]
\textit{\centering
\includegraphics[width = 0.96\columnwidth]{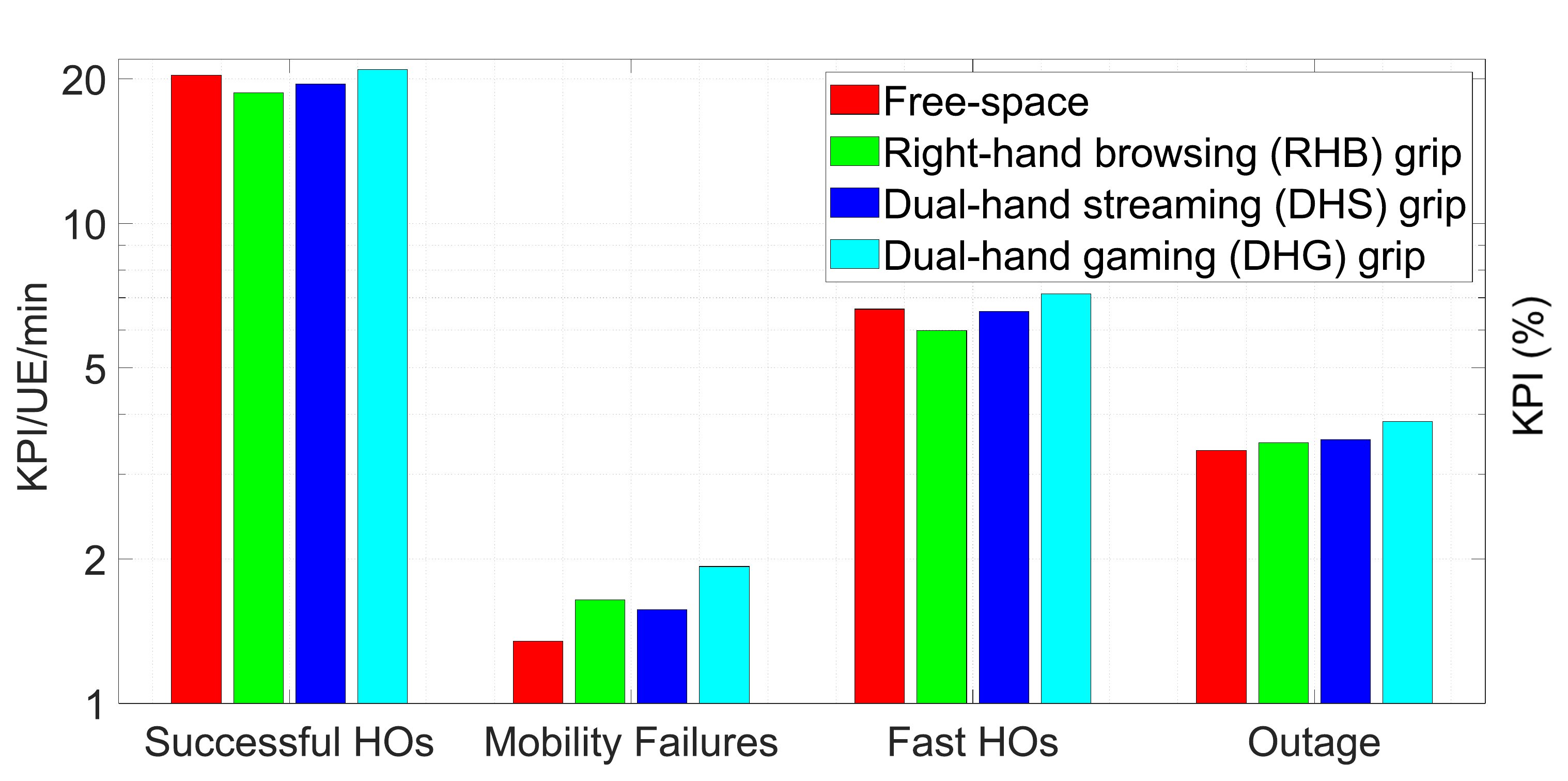}
\vspace{-0.2\baselineskip}  
\caption{A comparison of the mobility performance between the free-space case and the three hand blockage models.} 
\label{fig:Fig4}} 
\vspace{-1.1\baselineskip} \end{figure}

\begin{figure}[!t]
\textit{\centering
\includegraphics[width = 0.96\columnwidth]{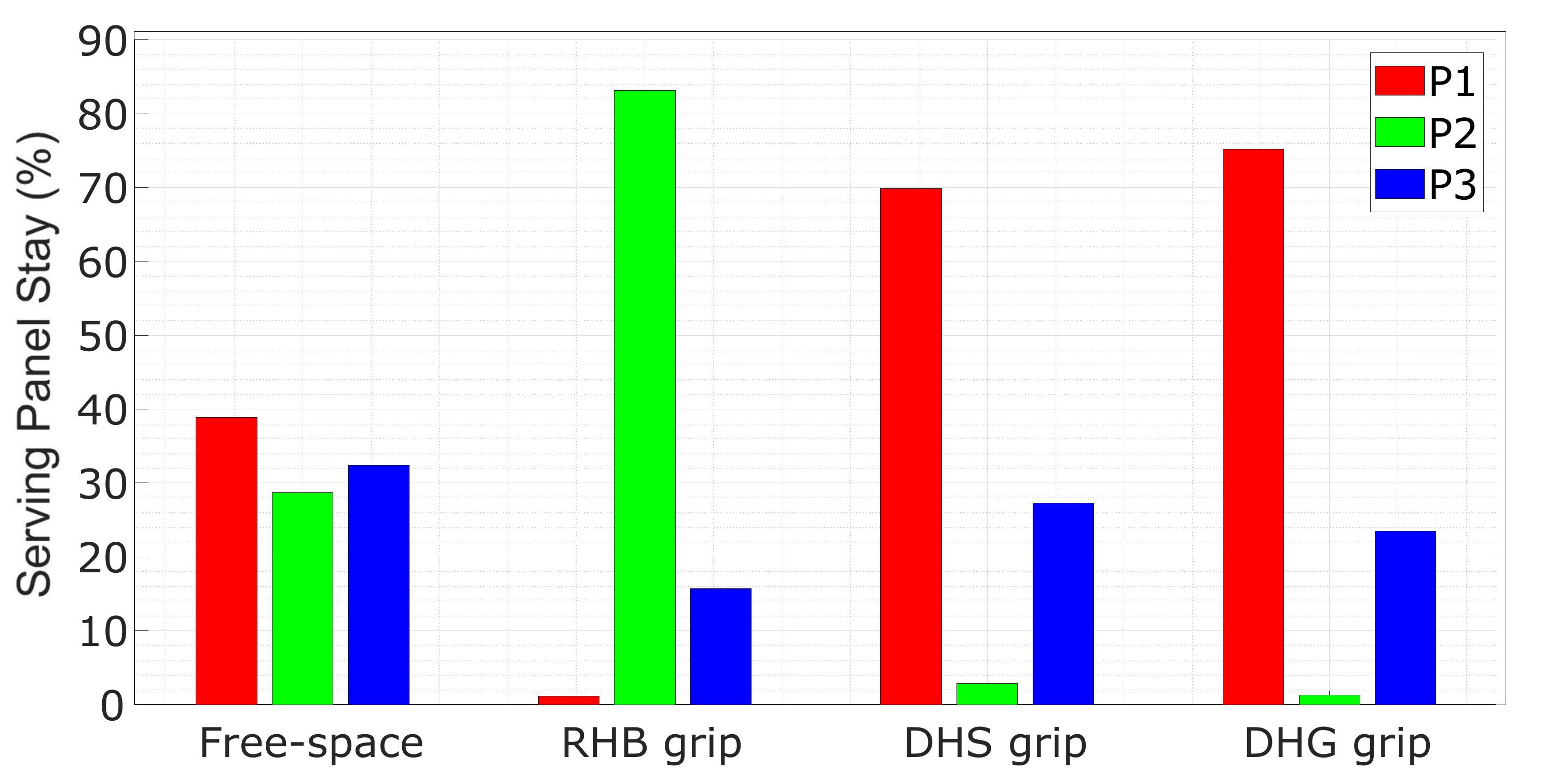}
\vspace{-0.4\baselineskip}  
\caption{A comparison of the serving panel stay percentage between the free-space case and the three different hand blockage models.} 
\label{fig:Fig2.5}} 
\vspace{-1.2\baselineskip} \end{figure}

Fig.\,\ref{fig:Fig4} shows a mobility performance comparison between the free-space case (no hand blockage) and the three different hand grips. When the RHB grip (shown in green) 
is compared with the free-space case (shown in red)
, it is seen that there is an approximate 22\% increase in mobility failures. This can also be reflected in the decrease in successful HOs and its subset fast HOs. Since it was mentioned in \Cref{Subection5.1} that the outage due to mobility failures is almost three times that of successful HOs, a relative increase of 4\% is observed in the outage. Interestingly, when the DHS grip 
is compared with the RHB grip, it is seen that the mobility failures reduce by 4.7\% and fast HOs increase by 9.5\%. The reduction in mobility failures primarily stems from the fact that P1 and P3 in the DHS grip (in Fig.\,\ref{fig:Fig2b}) now experience only partial blockage due to the hand positioning. This is illustrated by their increased usage in Fig.\,\ref{fig:Fig2.5}, which shows the serving panel usage for 
all blockage models.
Owing to the involvement of two hands in the DHS grip and the manner in which they completely encapsulate the hand, the hand reflection associated gains \cite[Section IV. B]{b6} are more pronounced, and consequently, fast HOs increase when compared with the RHB grip. When the DHG grip (shown in cyan) is compared with the DHS grip, it is seen that the mobility failures increase relatively by 23.0\% (43\% when compared with the free-space case), while fast HOs also increase by 8.7\%. Mobility failures increase because P2 in the DHG grip (in Fig.\,\ref{fig:Fig2c}) experiences complete blockage by the left hand. Additionally, owing to the hand positioning in the DHG grip, the hand reflection effect is now reduced and this also contributes to the increment in failures. This was also one of the main conclusions of \cite{b6}, where it was shown that hand reflections can be beneficial when one or more of the MPUE panels are not completely covered by the hand. Owing to the blockage on P2, the UE is now forced to use P1 and P3 even and the unreliable L3 RSRP measurements lead to more of such fast HOs \cite{b6}. This over-reliance on P1 and P3 is also clearly visible in Fig.\,\ref{fig:Fig2.5}. Since there is more outage now due to both more mobility failures and HOs, the DHG grip experiences the highest~outage.

\begin{figure}[!t]
\begin{subfigure}{0.49\textwidth}
\includegraphics[width=\textwidth]{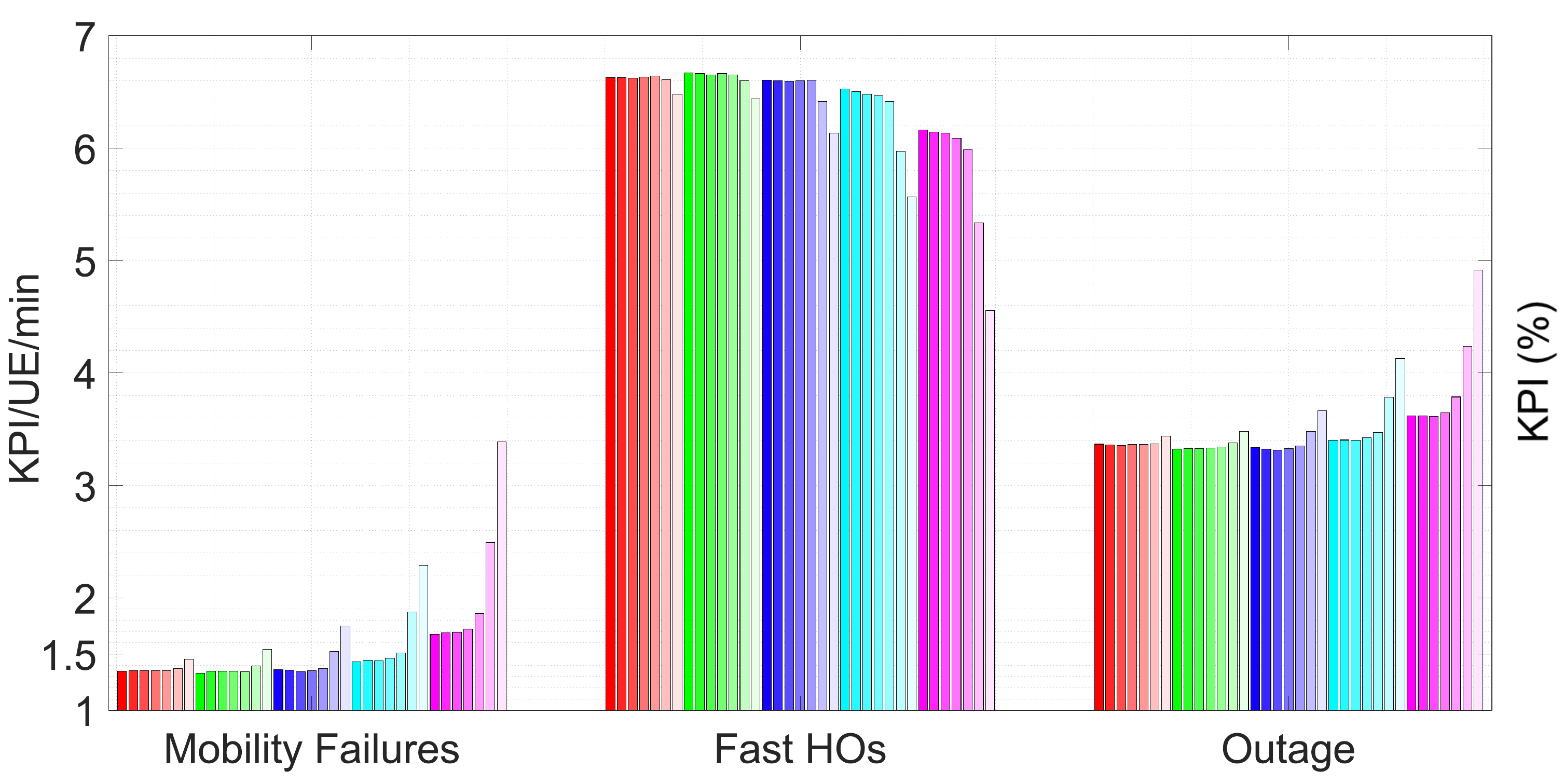}
\vspace{-2\baselineskip}
\label{fig:Fig5a}
\end{subfigure} \hfill   
  
\begin{subfigure}{0.5\textwidth}
\includegraphics[width=0.96\textwidth] {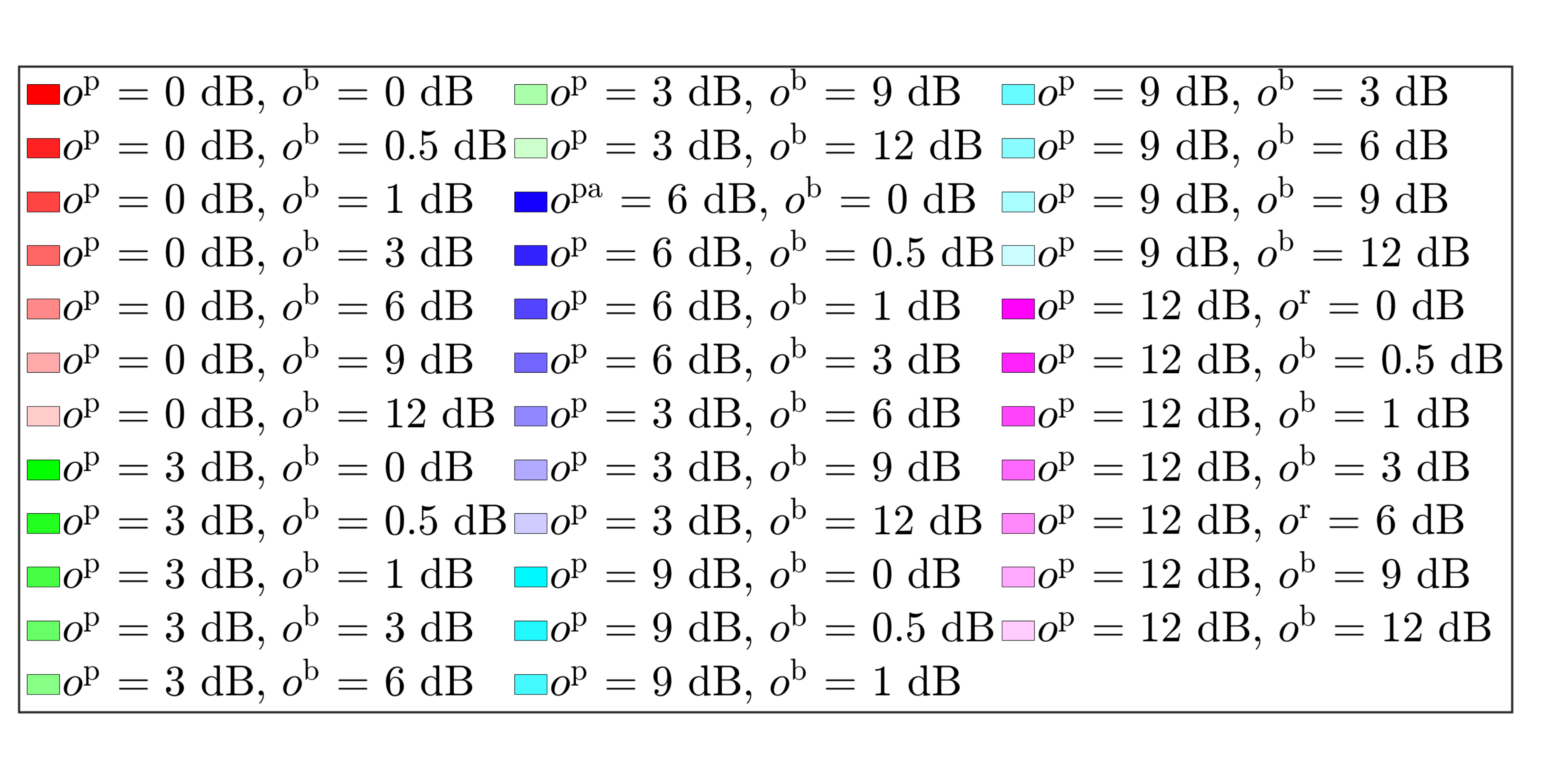}
\vspace{-1.1\baselineskip}
\caption{Mobility KPIs.}
\vspace{-1\baselineskip}
\label{fig:Fig5b}
\end{subfigure} \hfill   

\begin{subfigure}{0.5\textwidth}
\includegraphics[width=\textwidth]{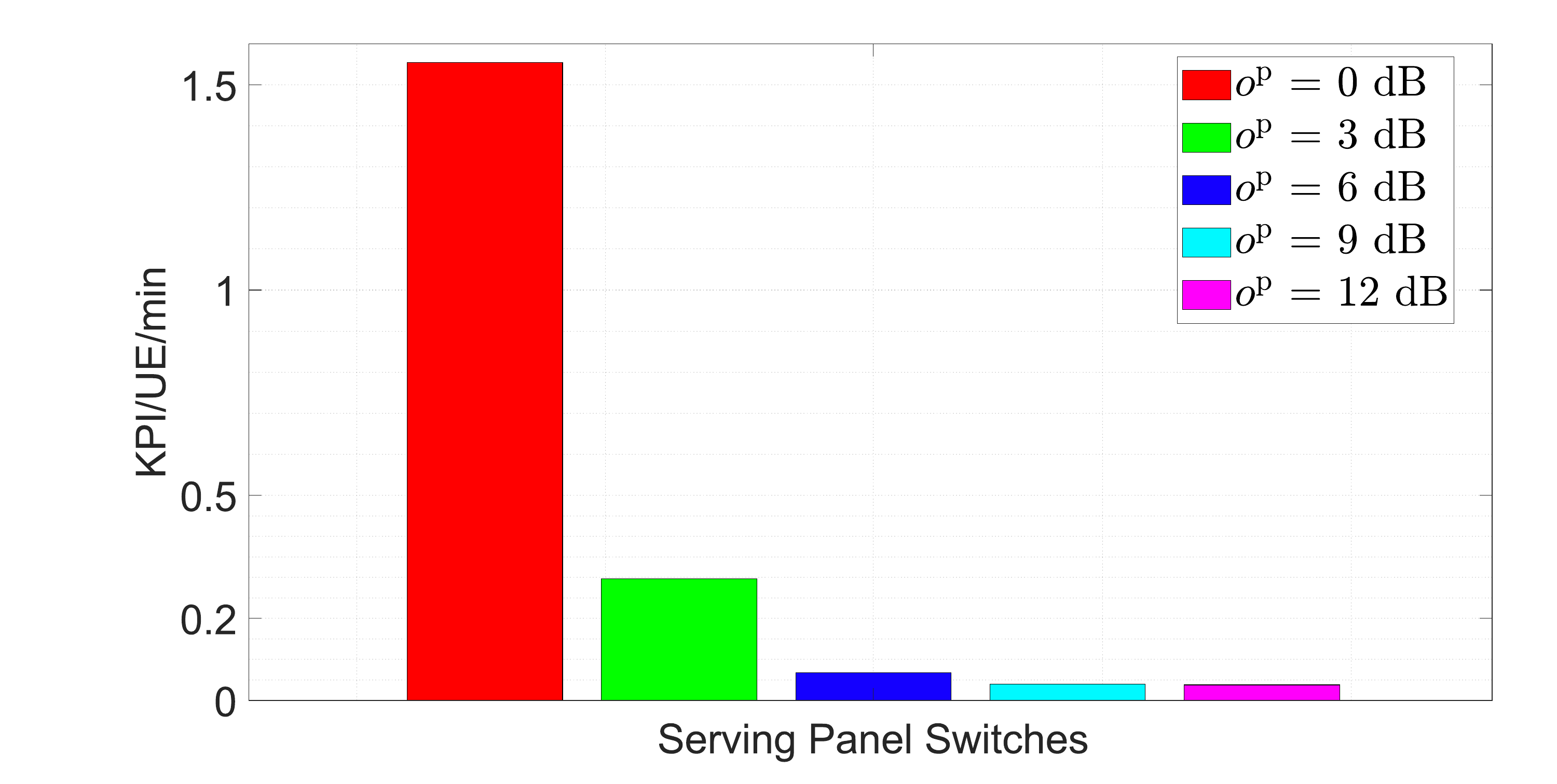}
\vspace{-1.6\baselineskip}
\caption{Serving panel switch frequency.}
\label{fig:Fig5c}
\end{subfigure}
  
\begin{subfigure}{0.5\textwidth}
\includegraphics[width=\textwidth] {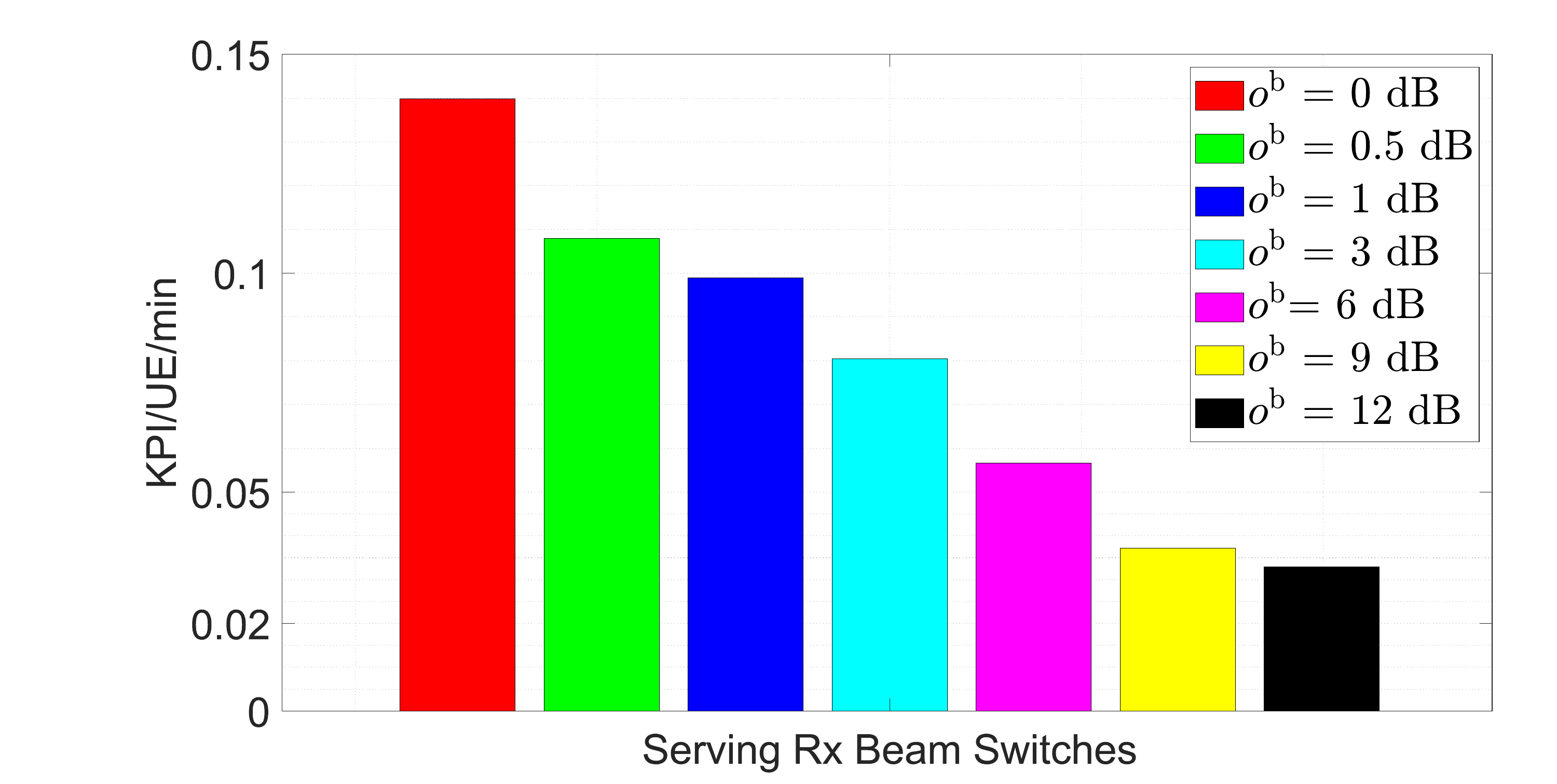}
\vspace{-1.4\baselineskip}
\caption{Serving Rx beam switch frequency.}
\label{fig:Fig5d}
\end{subfigure} 
  
\vspace{-0.2\baselineskip}  
\caption{A mobility performance comparison for different serving panel switching offset $o^\mathrm{p}$ and serving Rx beam switching offset $o^\mathrm{b}$ values.}
\label{fig:Fig5}
\vspace{-1.2\baselineskip} \end{figure}

In Fig.\,\ref{fig:Fig5} the mobility performance comparison for different $o^\mathrm{p}$ and $o^\mathrm{b}$ is depicted, where the RHB grip is considered. The performance comparison is made in terms of mobility KPIs and the panel and Rx beam switching frequencies, both of which impact UE power consumption as discussed in \Cref{Subection2.4}. In Fig.\,\ref{fig:Fig5b} it is seen that the mobility performance does not differ much for lower offsets, i.e., $o^\mathrm{p}
\leq \SI{6}{dB}$
and $o^\mathrm{b}
\leq \SI{3}{dB}$.
However, for higher values, i.e., $o^\mathrm{p}
\geq \SI{9}{dB}$
and $o^\mathrm{b}
\geq \SI{6}{dB}$,
it is observed that the mobility failures increase sharply while fast HOs decrease sharply. This is because in such cases persisting with the current serving panel or serving Rx beam is detrimental to the mobility performance and it would be better to make the switch, as determined by (\ref{Eq4}) and (\ref{Eq5}) to a better panel and Rx beam, respectively. The detrimental effect of these higher offset values is also clearly visible in the outage. In contrast, the lower $o^\mathrm{p}$ and $o^\mathrm{b}$ values are practical and do not degrade the mobility performance by a large margin. At the same time, it can be seen in Fig.\,\ref{fig:Fig5c} that the lower $o^\mathrm{p}$ values already yield a significant gain in terms of reducing the serving panel switches. For $o^\mathrm{p}  = \SI{6}{dB}$, the panel serving switches reduce by almost 96\% when compared to $o^\mathrm{p}  = \SI{0}{dB}$.  The same trend is seen for $o^\mathrm{b} = \SI{3}{dB}$ with a reduction of 42\%.

\section{Conclusion} \label{Section6}
In this paper, the mobility performance of three practical hand grips in terms of mobility key performance indicators is analyzed for MPUE, where beamforming is employed both on the gNb side and the UE side. It is well known that for multi-beam networks operating at FR2 the hand blockage effect is too considerable to be ignored. This paper is a novel attempt to understand how the mobility performance varies with different hand grips corresponding to different use cases, e.g., browsing, streaming, or gaming, where both the MPUE orientation and hand blockage effect involving one or both hands change. It is seen that the dual-hand gaming grip fares the worst, where mobility failures increase by 43\% as compared to the free-space case without any hand blockage. This must be taken into account by UE manufacturers in their devices based on the MPUE architecture to ensure seamless connectivity. Furthermore, a tradeoff between the mobility key performance indicators and panel and Rx beam switching frequency is also studied, where it is seen that both the panel and beam switches can be substantially reduced while hardly compromising on the mobility performance. This is important because both types of switches significantly impact UE power consumption. Based on this work, further studies to study the effect of Rx-beamforming with hand blockage combined with user-induced MPUE rotations~\cite{b5} can be carried out.

\vspace{-0.1\baselineskip}

\end{document}